\documentstyle[prl,aps,twocolumn,epsf,floats]{revtex}
\begin{document}
\draft
\twocolumn[
\hsize\textwidth\columnwidth\hsize\csname @twocolumnfalse\endcsname

\title{\bf Experiments on Critical Phenomena in a Noisy Exit Problem}

\author{D.~G. Luchinsky$^{(a),}$\cite{byline}, R.~S. Maier$^{(b,d)}$, 
R.~Mannella$^{(c)}$, P.~V.~E. McClintock$^{(a)}$, and D.~L. Stein$^{(d,b)}$}

\address{$^{(a)}$Department of Physics, Lancaster University, Lancaster
LA1 4YB, UK \protect\\
$^{(b)}$Department of Mathematics, University of Arizona, Tucson,
AZ 85721, USA \protect\\
$^{(c)}$Dipartimento di Fisica, Universit\`{a} di Pisa, Piazza Torricelli~2,
56100 Pisa, Italy \protect\\
$^{(d)}$Department of Physics, University of Arizona, Tucson, AZ 85721,
USA}

\date{July 1, 1997}
\maketitle
\widetext

\begin{abstract}

We consider noise-driven exit from a domain of attraction in a
two-dimensional bistable system lacking detailed balance. Through
analog and digital stochastic simulations, we~find a theoretically
predicted bifurcation of the most probable exit path as the parameters
of the system are changed, and a corresponding nonanalyticity of the
activation energy. We~also investigate the extent to which the
bifurcation is related to the local breaking of time-reversal
invariance. 

\end{abstract}

]
\narrowtext

\noindent Noise-induced motion away from a locally stable state, in a
system far from thermal equilibrium, arises in diverse scientific
contexts, e.g., glasses~\cite{glasses}, arrays of Josephson
junctions~\cite{kautz}, stochastically modeled computer
networks~\cite{network}, stochastic resonance~\cite{sr}, and stochastic
ratchets~\cite{magnasco}. Because these systems in~general lack
detailed balance, progress in understanding this phenomenon has been
slower than in thermal equilibrium systems. In~particular, there exist
no simple or general relations from which the rate of noise-induced
transitions between stable states can be obtained.

Recently, substantial progress on the nonequilibrium case has been
achieved in the limit of weak noise, using path integral or equivalent
Hamiltonian formulations
\cite{bray89,McKane95,ms93-I,dms-Smel97}. 
Fluctuational motion
of the system can then be characterised by the pattern of
optimal (i.e.~most probable) fluctuational trajectories. An~optimal
trajectory is one along which a system moves, with overwhelming
probability, when it fluctuates away from a stable state toward a
specified remote state. These are rare events but, when they occur,
they do~so in an almost deterministic way: e.g.\ escape from a domain of
attraction typically follows a {\it unique trajectory\/}. The
properties of this most probable exit path (MPEP) determine the
weak-noise behaviour of the mean first passage time (MFPT).

In recent years, it has been realized that in nonequilibrium systems,
the pattern of optimal fluctuational trajectories may contain
``focusing singularities''
\cite{dms-Smel97,jauslin}. 
Their
effect on exit phenomena was considered by Maier and Stein~\cite{ms93,ms96}
who showed that, for a symmetric double well system (see below), the MPEP
bifurcates when the model parameters are changed in such a way that a
focusing singularity appears along~it. That~is, the MPEP ceases to be
unique. This bifurcation breaks the symmetry of the model, and is
accompanied by a nonanalyticity of the activation energy for inter-well
transitions: it~is analogous to a second-order phase transition in a
condensed matter system~\cite{ms96}. This analogy throws new light~on,
e.g., exit bifurcation phenomena in systems driven by coloured
noise~\cite{McKane95}.

Many of these theoretical ideas, although important, remain untested
experimentally or numerically.  In~this Letter we use an analog experiment
and numerical simulations to demonstrate the predicted bifurcation of the
MPEP\null, and the corresponding nonanalytic behavior of the activation
energy and related quantities.  We~investigate the nature of the broken
symmetry in detail, and show how bifurcation is accompanied by a loss of
time-reversal invariance along the MPEP\null.

We investigate the motion of an overdamped particle in the
two-dimensional drift field first proposed in~\cite{ms93-I}: $ {\bf
u}(x,y) = (x-x^3-\alpha xy^2, -y-x^2y)$, where $\alpha$~is a parameter.
It has point attractors at $(\pm 1,0)$ and a saddle point at $(0,0)$.
If~the particle is subject to additive isotropic white noise ${\bf
f}(t) = (f_x, f_y)$, its position $(x,y)$ will satisfy the coupled
Langevin equations

\begin{eqnarray}
&&\dot x =x-x^3-\alpha xy^2 + f_x(t), \nonumber\\ 
&&\dot y = -y- x^2y + f_y(t), \label{system}\\[\jot]
&&\langle f_i(t) \rangle = 0,\quad \langle f_i(s)f_j(t)
\rangle = D\delta_{ij}\delta(s-t) \nonumber.
\end{eqnarray}

\noindent Since ${\bf u}$~is not a gradient field (unless $\alpha =
1$), the dynamics will not satisfy detailed balance. The Fokker--Planck
equation for the particle's probability density $\rho=\rho(x,y,t)$
will be

\begin{equation}
\label{fpe}
\dot \rho =(D/2)\nabla ^2\rho -{\bf \nabla \cdot }(\rho{\bf u}).
\end{equation}

In the weak-noise limit, escape of the particle from the domain of
attraction of either fixed point $(x_s,0)=(\pm1,0)$ is governed by the the
slowest-decaying nonstationary eigenmode of the Fokker--Planck
operator~\cite{carmeli91}, whose eigenvalue $\lambda_1$ becomes
exponentially small as ${D\to0}$.  In~this limit the MFPT $\langle t_{\rm
exit}\rangle$ is well approximated by~$\lambda_1^{-1}$. The
slowest-decaying eigenmode is called the {\it quasistationary probability
density\/}; we~denote it by~$\rho_1$. It~may be approximated in a WKB-like
fashion~\cite{ms93-I,ms93,ms96}, i.e.,

\begin{equation}
\label{wkb}
\rho_1(x,y)\sim K(x,y) \exp\left(-W(x,y)/D\right), \qquad D\to0.
\end{equation}

\noindent
Here $W(x,y)$ may be viewed as a {\it classical action at zero energy\/},
since it turns~out to satisfy an eikonal (Hamilton--Jacobi) equation of the
form $H({\bf x}, {\bf \nabla}W) = 0$, where $ H({\bf x}, {\bf p}) = \frac12
{\bf p}^2 + {\bf u}({\bf x}) \cdot {\bf p}$ is a so-called
Wentzel--Freidlin Hamiltonian~\cite{freidlin}. The optimal fluctuational
trajectories are projections onto coordinate space of the zero-energy
classical trajectories determined by this Hamiltonian.  These lie on the
three-dimensional energy surface specified by~${H=0}$, embedded in the
four-dimensional phase space with coordinates $(x,y,p_x,p_y)$.  In~general,
the computation of~$W(x,y)$ requires a minimisation over the set of
zero-energy trajectories starting at~$(x_s,0)$ and terminating
at~$(x,y)$. Moreover the MPEP\null, for the domain of attraction
of~$(x_s,0)$, is the zero-energy trajectory of least action which extends
from~$(x_s,0)$ to the saddle~$(0,0)$. The MPEP action $\delta W\equiv
W(0,0)-W(x_s,0)$ governs the weak-noise behavior of the MFPT\null.
To~leading order it is of the activation type, i.e.,

\begin{equation}
\langle t_{\rm exit} \rangle \sim {\rm const} \times 
e^{\delta W / D}, \qquad D\to0.
\end{equation}

\noindent So $\delta W$ is interpreted as an activation energy. The
prefactor `const' is determined by the function $K(x,y)$ of~(\ref{wkb}). 

When $\alpha=1$, the dynamics of the particle satisfy detailed balance,
and the pattern of optimal trajectories emanating from $(x_s,0)$
contains no singularities. It~was found earlier \cite{ms93,ms96} that,
as $\alpha$~is increased, the first focusing singularity on the MPEP
(initially lying along the $x$-axis) appears when $\alpha =\alpha_c
\equiv 4$. It~signals the appearance of a transverse `soft mode,' or
instability, which causes the MPEP to bifurcate. Its physical origin is
clear: as~$\alpha$~is increased, the drift toward $(x_s,0)$ `softens'
away from the $x$-axis, which eventually causes the on-axis MPEP to
split. The two new MPEP's move off-axis, causing the activation energy
(previously constant) to start decreasing. So~the activation energy as
a function of~$\alpha$ is nonanalytic at $\alpha=\alpha_c$. If~$\alpha$
is increased substantially beyond~$\alpha_c$, further bifurcations of
the on-axis zero-energy classical trajectory occur when $\alpha$~equals
$\alpha_c^{(j)} \equiv (j+1)^2$, where $j$~is the number of the
bifurcation. But the oscillatory trajectories arising from such
bifurcations are believed to be unphysical, since the
on-axis trajectory is no~longer the MPEP\null.  (Cf.~\cite{dms-Smel97}.)

To test these theoretical predictions, and to seek further insight into the
nature of the broken symmetry, we have built an analog electronic model of
the system~(\ref{system}) using standard
techniques~\cite{mccmoss}. We~drive it with zero-mean quasi-white Gaussian
noise from a noise-generator, digitize the response $x(t), y(t)$, and
analyse it with a digital data processor. Transition probabilities are
measured by a standard level-crossing technique.  Experimental
investigations of the optimal fluctuational trajectories are based on
measurements of the prehistory probability
distribution~\cite{Dyk92,luch96}. This method was recently extended to
include analysis of relaxational trajectories and thus to investigate
directly the presence or absence of time-reversal symmetry and detailed
balance~\cite{luch-jpa-irrev}.

We have also carried out a complementary digital simulation
of~(\ref{system}) using the algorithm of~\cite{algo}, with particular
attention paid to the design of the noise-generator on~account of the long
simulation times. Transition probabilities were measured using a
well-to-well method, and the analysis of the data to extract the optimal
fluctuational and relaxational trajectories was based on a method similar
to that used in the analog experiments.

\begin{figure}
\centering
\epsfxsize=2.95in
\leavevmode
\epsfbox{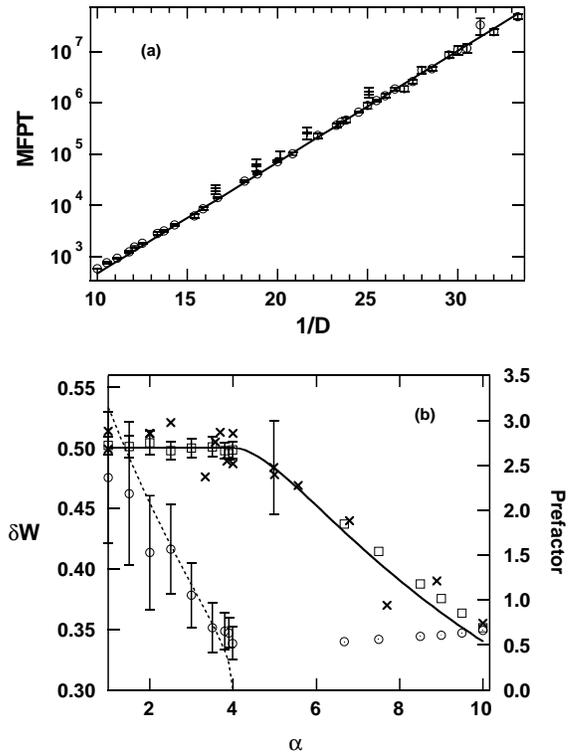}
\caption{(a)~The mean first passage time as a function of
inverse noise intensity $1/D$ for $\alpha = 1$, from analog experiment
(bars), numerical simulation (circles) and calculation (solid line).
(b)~The inter-well activation barrier~$\delta W$, as~a function
of~$\alpha$, from analog experiment (crosses), numerical simulation
(squares) and theory~\protect\cite{ms93,ms96} (full line). The dashed curve
and circle data represent the MFPT prefactor from calculation and numerical
simulation respectively.} 
\end{figure}

Some activation energy results are shown in Fig.~1. Part~(a) plots the
MFPT $\langle t_{\rm exit}\rangle$ as a function of inverse noise
intensity~$1/D$ for the special case $\alpha = 1$. In~this case the
drift field is the gradient of the potential $U(x,y) = \bigl(y^2(1+x^2)
- x^2 +x^4/2\bigr)/2$, and $W$~can be obtained exactly ($W=2U$). The
analog and digital results are in good agreement, and demonstrate that
the noise dependence of the MFPT is indeed of the activation type
predicted by the theory. Activation energies determined from the slopes
of a series of plots like those in Fig.~1(a), yielded the results shown
in Fig.~1(b), where they are compared with theoretical values
of~$\delta W$ determined from the true (least action) MPEP or
MPEP's~\cite{ms93,ms96}. At the predicted critical value $\alpha_c =
4$, marked changes in both the activation energy and MFPT prefactor
(which $\rightarrow 0$) are evident: theory predicts that the
activation energy bifurcates here into two values, corresponding to
paths on and off the $x$-axis, of which only the latter (lower action)
path is expected to be physically meaningful. The dependence of the
activation energy on~$\alpha$ near the second critical value
$\alpha_c^{(2)} \equiv 9$ is smooth, in agreement with the prediction
that higher bifurcations correspond to folding of a nonphysical sheet
of the `action surface' $W=W(x,y)$, and are not
observable \cite{dms-Smel97,ms96}.

Interestingly, the transition shown in Fig.~1(b) resembles the
bifurcation of the activation energy in an overdamped oscillator driven
by coloured noise~\cite{McKane95}. This suggests that the WKB
analysis~\cite{ms93,ms96} of~(\ref{system}) may provide physical and
topological insight into the corresponding transition phenomena in
systems driven by quasi-monochromatic noise.

\begin{figure} 
\centering
{\leavevmode\epsfxsize=2.95in\epsfbox{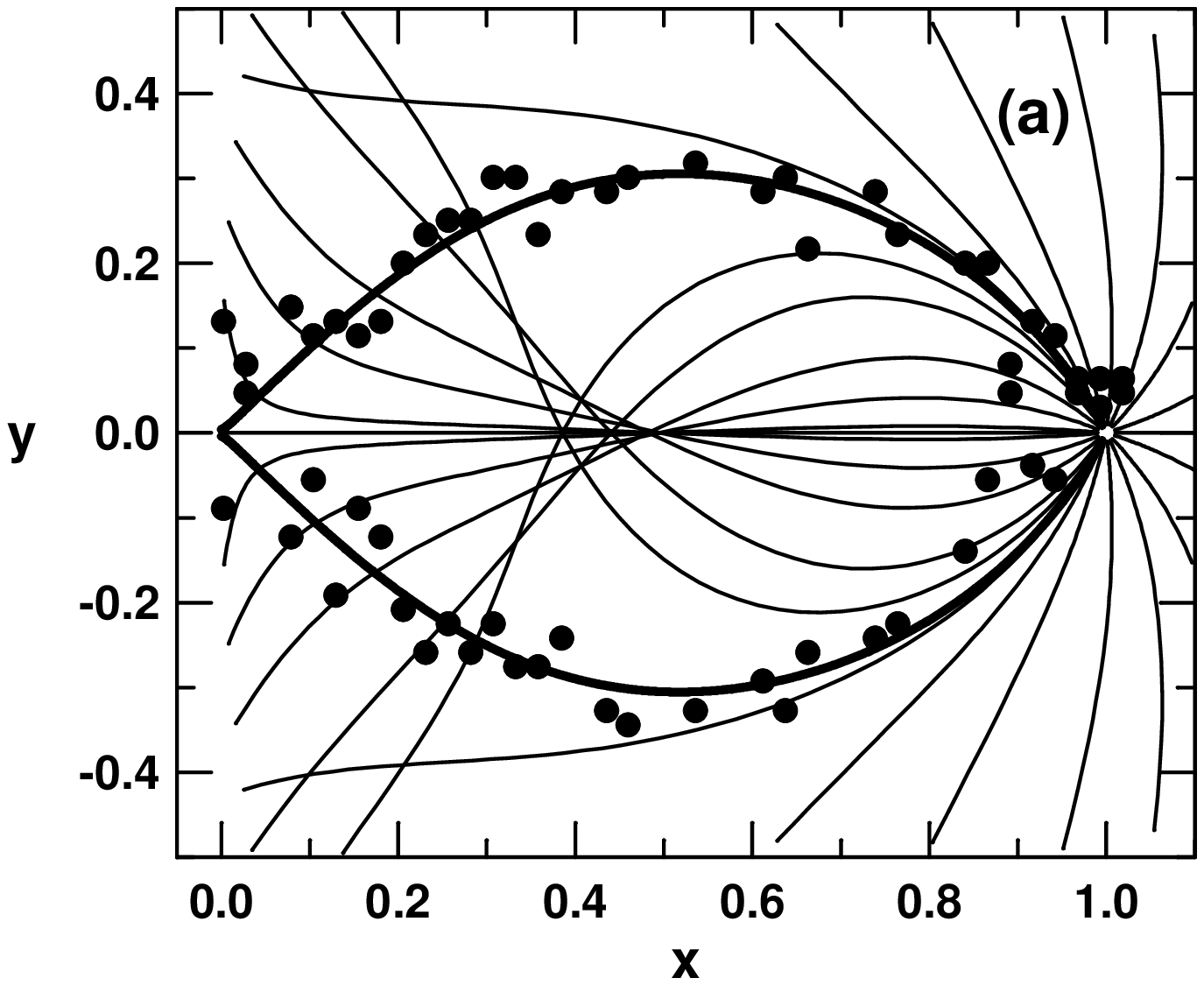}}
{\leavevmode\epsfxsize=2.95in\epsfbox{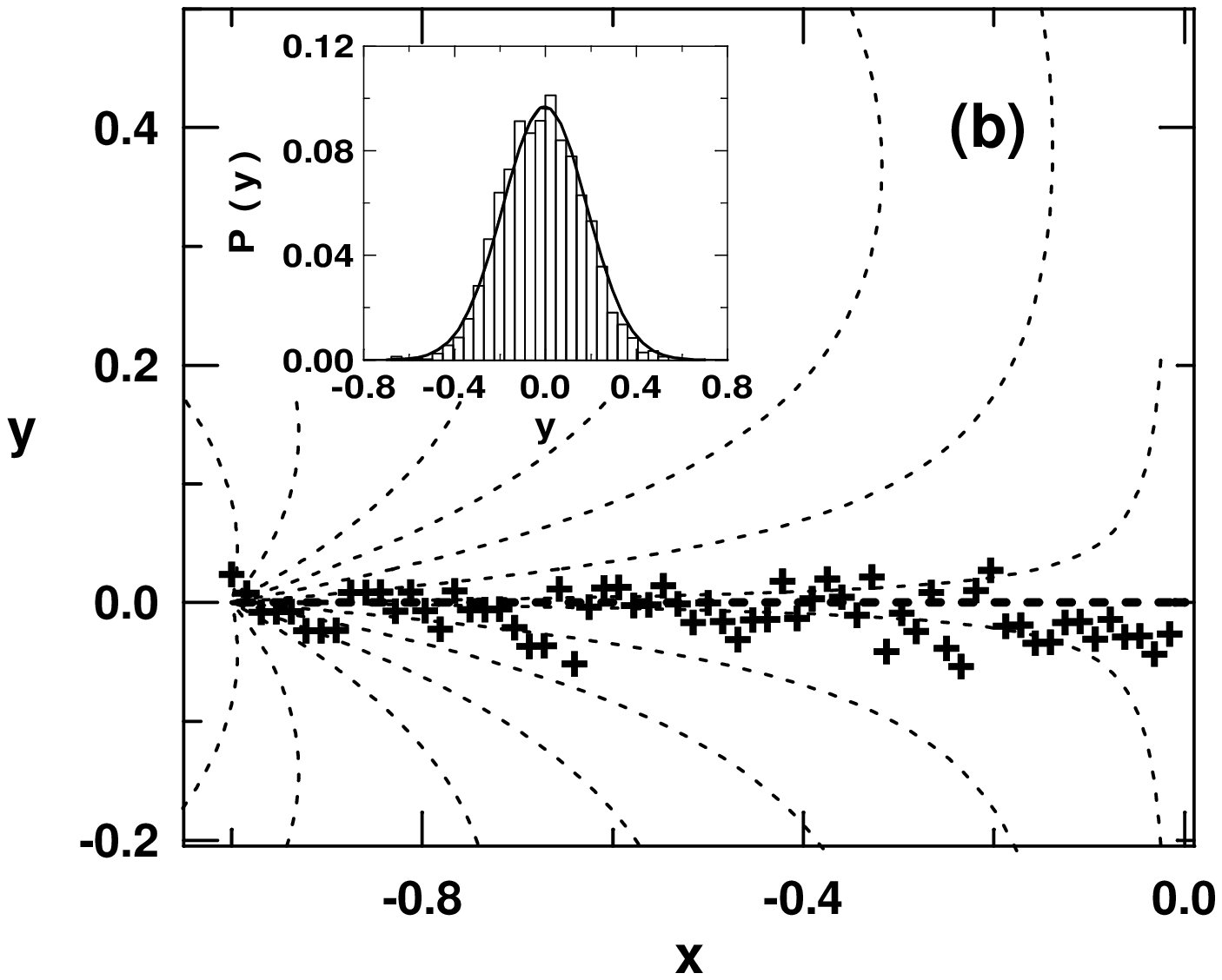}}
\caption{Measured positions of the ridges (first moments) for
$\alpha = 6.67$ of (a)~the fluctuational part (filled circles) and (b)~the
relaxational part (pluses) of the escape trajectories. Nearby theoretical
trajectories are shown by lines. The insert shows the exit location
distribution along the $y$-axis.} 
\end{figure}

To verify experimentally the expected relationship between the
bifurcation of the MPEP and the bifurcation of the activation energy,
we have measured two-dimensional prehistory probability distributions
\cite{Dyk92} of fluctuational trajectories bringing the system into the
vicinity of the separatrix between the two wells (the $y$-axis).
In~the limit of low noise intensity, the maxima of the corresponding
distributions trace out optimal trajectories
\cite{luch96,luch-jpa-irrev}. The positions of these maxima are
compared to the calculated MPEP's for $\alpha = 6.67$ in Fig.~2(a).
It~is clear that the typical fluctuational path corresponding to escape
from the domain of attraction of $(x_s,0)$ follows very closely one of
the predicted MPEP's.

To seek further experimental insight into the character of the broken
symmetry for the MPEP\null, we have also followed the dynamics of the
relaxational part of the escape paths, after they have crossed the $y$-axis
separatrix. The prehistory and relaxational probability distributions
provide a complete history of the time evolution of large fluctuations to
and from a given remote state.  One can thus investigate experimentally
detailed balance and time-symmetry (or~the lack of
them)~\cite{luch-jpa-irrev}. The positions of the maxima of the measured
relaxational distributions are compared with the corresponding theoretical
trajectories in Fig.~2(b). A~detailed analysis of the distributions will be
given elsewhere.  It~can be seen from the figure that for $\alpha >
\alpha_c$ the MPEP breaks time-reversal symmetry, i.e., the average growth
and average decay of fluctuations~\cite{Onsager31} traced out by the ridges
of the corresponding distributions take place along trajectories that are
asymmetric in time. That~is, for $\alpha > \alpha_c$ the MPEP is {\it
not\/} a time-reversed relaxational trajectory.

The inset in Fig.~2(b) shows the distribution of points where the escape
trajectories hit the $y$-axis separatrix (i.e.,~the exit location
distribution).  Its shape is nearly Gaussian, as expected from the saddle
point approximation of~\cite{ms96}. The maximum is situated near the saddle
point clearly demonstrating that, in the limit of weak noise, exit occurs
via the saddle point.

The relationship between time-reversal symmetry-breaking for the MPEP
when ${\alpha > \alpha_c}$, and symmetry-breaking generally for the
system~(\ref{system}), is quite subtle. The system loses detailed
balance and time reversal symmetry as soon as $\alpha > 1$~ and the
drift field~$\bf u$ becomes nongradient. It~is on~account of a special
symmetry of the system (reflection symmetry through the $x$-axis) that
the MPEP can remain unchanged in this nongradient drift field up~to the
value $\alpha_c=4$. Thus, for $1 < \alpha < 4$ the dynamics of the most
probable fluctuational trajectories is a mirror-image of the
relaxational dynamics {\it only\/} along the $x$-axis; everywhere else
in the domain of attraction of $(x_s,0)$ the outward optimal
trajectories are not antiparallel to the inward relaxational
trajectories, and the resulting closed loops enclose nonzero
area~\cite{ms93,freidlin}.

\begin{figure}
\centering
{\leavevmode\epsfxsize=2.5in\epsfbox{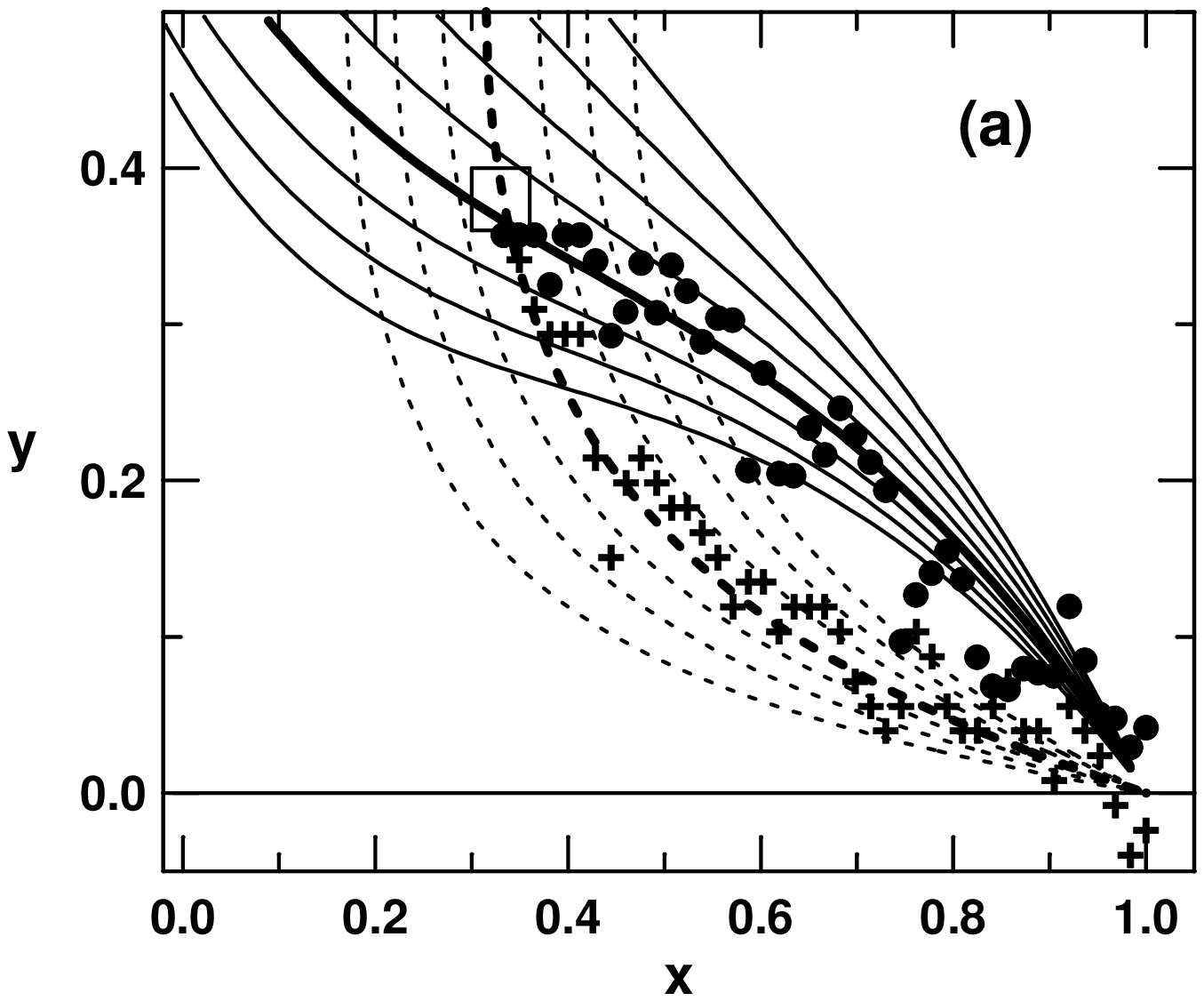}}
{\leavevmode\epsfxsize=2.5in\epsfbox{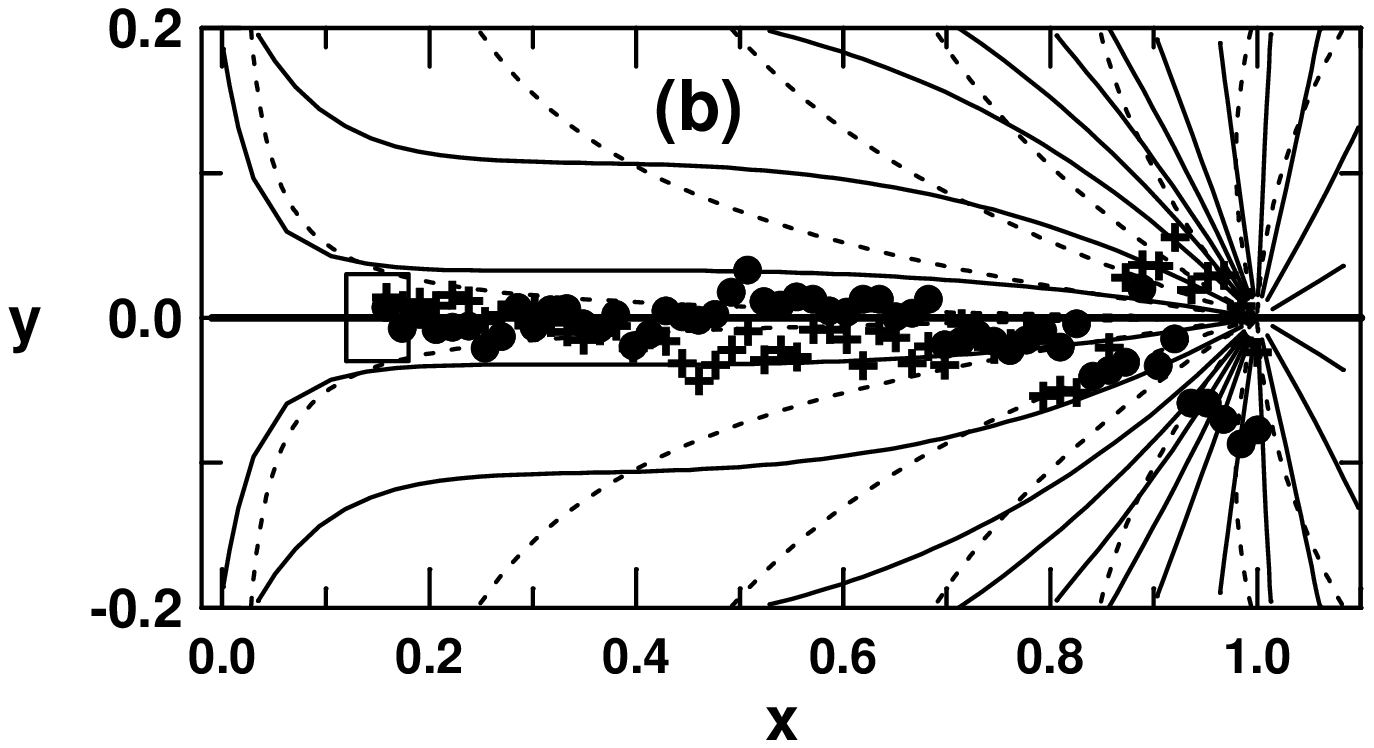}}
\caption{Demonstration of local properties of detailed balance and
time-reversal symmetry of~(\protect\ref{system}) for $\alpha =
3.5$.  (a)~Measured positions of the ridges of the fluctuational (filled
circles) and relaxational (pluses) parts of the trajectories from
$(1,0)$ to $(0.3, 0.3)$, compared with theoretical predictions
\protect\cite{ms93,ms96} (curves).  (b)~The same for trajectories
extending to the on-axis remote state $(0.1, 0)$.}
\end{figure}

This prediction has been tested experimentally by tracing out optimal
paths to/from specified remote states both on and off the $x$-axis, for
$1 < \alpha < \alpha_c$. Some results are shown in Fig.~3 for $\alpha =
3.5$. It is evident that the ridges of the fluctuational (filled
circles) and relaxational (pluses) distributions follow closely the
theoretical curves. For an off-axis remote state (Fig.~3(a)), they form
closed loops of nonzero area, thus demonstrating the expected
rotational flow of the probability current in a nonequilibrium
system~\cite{Onsager31}. The corresponding ridges for an on-axis remote
state (Fig.~3(b)) are antiparallel, indicating that symmetry is
preserved along the $x$-axis.

Our results verify the predicted bifurcation of the MPEP in
(\ref{system}) at $\alpha = \alpha_c \equiv 4$, with a corresponding
nonanalyticity of the activation energy. We~have demonstrated that, in
the limit $D \to 0$, detailed balance and time-reversal symmetry can be
considered as local properties along the MPEP of the system in the
sense discussed above, and that the bifurcation phenomenon can be
related to local time-reversal symmetry-breaking along the MPEP\null:
results that may bear on two-dimensional stochastic ratchets
\cite{Sym-rat} where symmetry plays an important role.
Having thus demonstrated (see also~\cite{luch96}) the reality of
phenomena inferred from $D \to 0$ optimal paths, we~anticipate that
other important $D \to 0$ theoretical predictions, e.g.\ ``cycling'' of
the exit location distribution~\cite{ms96-I}, will also be physically
realisable.

The research was supported by the Engineering and Physical Sciences
Research Council (UK), the Royal Society of London, the Russian
Foundation for Basic Research, the National Science Foundation~(US), and
the Department of Energy~(US).



\begin{thebibliography}{99}

\bibitem[*]{byline} Permanent address: Institute of Metrological Service,
Ozernaya~46, Moscow 119361, Russia.

\bibitem{glasses} D.~L. Stein, R.~G. Palmer, J.~L. van Hemmen, and
C.~R. Doering, Phys.\ Lett.~A {\bf 136}, 353 (1989).

\bibitem{kautz} R.~L. Kautz, Rep.\ Progr.\ Phys.\ {\bf 59}, 935
(1996).

\bibitem{network} R.~S. Maier, in {\it Proc.\ 33rd Annual Allerton
Conference on Communication, Control, and Computing\/} (Monticello,
Illinois, Oct.\ 1995), 766.

\bibitem{sr} See special issue of Nuovo Cim. D {\bf 17}, nos.\ 7--8
(1995); A.~R. Bulsara and L.~Gammaitoni, Phys.\ Today {\bf 49},
no.~3, 39 (1996).

\bibitem{magnasco} M. Magnasco, Phys.\ Rev.\ Lett.\ {\bf 71}, 1477
(1993).


\bibitem{bray89}  A.~J. Bray and A.~J. McKane, Phys.\ Rev.\ Lett.\ {\bf 62},
493 (1989).

\bibitem{McKane95} S.~J.~B. Einchcomb and A.~J. McKane, 
Phys.\ Rev.~E {\bf 51}, 2974 (1995).


\bibitem{ms93-I} R.~S. Maier and D.~L. Stein, Phys.\ Rev.~E {\bf 48},
931 (1993).

\bibitem{dms-Smel97} M.~I. Dykman, M.~M. Millonas, and V.~N. Smelyanskiy, 
Phys.\ Lett.\ A {\bf 195}, 53 (1994), cond-mat/9410056; 
V.~N. Smelyanskiy, M.~I. Dykman, and
R.~S. Maier, Phys.\ Rev.~E {\bf 55}, 2369 (1997).


\bibitem{jauslin} H.~R. Jauslin, Physica {\bf 144A}, 179 (1987);
M.~V. Day, Stochastics {\bf 20}, 121 (1987).

\bibitem{ms93} R.~S. Maier and D.~L. Stein, Phys.\ Rev.\ Lett.\ {\bf
71}, 1783 (1993).

\bibitem{ms96} R.~S. Maier and D.~L. Stein, J.~Stat.\ Phys.\ {\bf
83}, 291 (1996), cond-mat/9506097.

\bibitem{carmeli91} B.~Carmeli, V.~Mujica, and A.~Nitzan, Berichte der
Bunsen-Gesellschaft {\bf 95}, 319 (1991).

\bibitem{freidlin} M.~I. Freidlin and A.~D. Wentzell, {\it Random
Perturbations of Dynamical Systems\/} (Springer-Verlag, New York/Berlin,
1984).

\bibitem{mccmoss} L.~Fronzoni, in {\it Noise in Nonlinear Dynamical
Systems\/}, edited by F.~Moss and P.~V.~E. McClintock (Cambridge University
Press, Cambridge, England, 1989), vol.~3, 222; P.~V.~E. McClintock and
F.~Moss, {\it op.~cit.}, 243.

\bibitem{Dyk92} M.~I. Dykman, P.~V.~E. McClintock, V.~N. Smelyanskiy, N.~D.
Stein, and N.~G. Stocks, Phys.\ Rev.\ Lett.\ {\bf 68}, 2718 (1992).

\bibitem{luch96} M.~I. Dykman, D.~G. Luchinsky, P.~V.~E. McClintock, and
V.~N. Smelyanskiy, Phys.\ Rev.\ Lett.\ {\bf 77}, 5229 (1996).

\bibitem{luch-jpa-irrev} D.~G. Luchinsky, ``On the nature of large fluctuations
in equilibrium systems: Observation of an optimal force''; 
D.~G. Luchinsky and P.~V.~E. McClintock, ``Irreversibility of classical
fluctuations.''  To~be published.

\bibitem{algo} R.~Mannella, ``Numerical integration of stochastic
differential equations,'' in {\it Proc.\ Euroconference on
Supercomputation in Nonlinear and Disordered Systems\/} (World
Scientific, Singapore, in~press).

\bibitem{Onsager31} L.~Onsager, Phys.\ Rev.\ {\bf 37}, 405 (1931).

\bibitem{Sym-rat} G.~W. Slater, H.-L. Guo, and G.~I. Nixon, 
Phys.\ Rev.\ Lett.\ {\bf 78}, 1170 (1997).

\bibitem{ms96-I} M.~V. Day, Stochastics {\bf 48}, 227 (1994);
J.~Dynamics and Differential Equations {\bf 8}, 573 (1996);
R.~S. Maier and D.~L. Stein, Phys.\ Rev.\ Lett.\ {\bf 77}, 4860 (1996),
cond-mat/9609075.

\end{thebibliography}
\end{document}